\documentclass[times,twocolumn,final]{elsarticle}

\usepackage{medima}
\usepackage{framed,multirow}

\usepackage{amssymb}
\usepackage{latexsym}

\usepackage{url}
\usepackage{xcolor}
\usepackage{color}

\usepackage{hyperref}

\usepackage{threeparttable}
\usepackage{amsmath,graphicx,booktabs}

\usepackage[font=small, labelfont=bf, labelsep=none]{caption}
\definecolor{red}{rgb}{.8,.349,.1}

\journal{Medical Image Analysis}

\begin{document}
	
	\verso{Jiamin Liang and Xin Yang \textit{et~al.}}
	
	\begin{frontmatter}
	    \setlength{\skip\footins}{0.5cm}
		
		\title{Sketch guided and progressive growing GAN for realistic and editable ultrasound image synthesis}%
		
		\author[1,2,3]{Jiamin \snm{Liang}\fnref{fn1}}
		\author[1,2,3]{Xin \snm{Yang}\fnref{fn1}}
		\fntext[fn1]{The two authors contribute equally to this work.}
		\author[1,2,3]{Yuhao \snm{Huang}}
		\author[1,2,3]{Haoming \snm{Li}}
		\author[1,2,3]{Shuangchi \snm{He}}
		\author[2,4]{Xindi \snm{Hu}}
		\author[1,2,3]{Zejian \snm{Chen}}
		\author[1,2,3]{\\Wufeng \snm{Xue}}
		\author[1,2,3]{Jun \snm{Cheng}\corref{cor1}}
		\ead{chengjun583@qq.com}
		\author[1,2,3]{Dong \snm{Ni}\corref{cor1}}
		\ead{nidong@szu.edu.cn}
		\cortext[cor1]{Corresponding author.}
		
		\address[1]{National-Regional Key Technology Engineering Laboratory for Medical Ultrasound, School of Biomedical Engineering, Health Science Center, Shenzhen University, Shenzhen, China}
		\address[2]{Medical Ultrasound Image Computing (MUSIC) Lab, Shenzhen University, Shenzhen, China}
		\address[3]{Marshall Laboratory of Biomedical Engineering, Shenzhen University, Shenzhen, China}
		\address[4]{School of Biomedical Engineering and Information, Nanjing Medical University, Nanjing, China}
		
		\received{17 June 2021}
		\accepted{13 April 2022}
		\availableonline{26 April 2022}

		\begin{abstract}
		Ultrasound (US) imaging is widely used for anatomical structure inspection in clinical diagnosis. The training of new sonographers and deep learning based algorithms for US image analysis usually requires a large amount of data. However, obtaining and labeling large-scale US imaging data are not easy tasks, especially for diseases with low incidence. Realistic US image synthesis can alleviate this problem to a great extent. In this paper, we propose a generative adversarial network (GAN) based image synthesis framework. Our main contributions include: 1) we present the first work that can synthesize realistic B-mode US images with high-resolution and customized texture editing features; 2) to enhance structural details of generated images, we propose to introduce auxiliary sketch guidance into a conditional GAN. We superpose the edge sketch onto the object mask and use the composite mask as the network input; 3) to generate high-resolution US images, we adopt a progressive training strategy to gradually generate high-resolution images from low-resolution images. In addition, a feature loss is proposed to minimize the difference of high-level features between the generated and real images, which further improves the quality of generated images; 4) the proposed US image synthesis method is quite universal and can also be generalized to the US images of other anatomical structures besides the three ones tested in our study (lung, hip joint, and ovary); 5) extensive experiments on three large US image datasets are conducted to validate our method. Ablation studies, customized texture editing, user studies, and segmentation tests demonstrate promising results of our method in synthesizing realistic US images.
			
		\end{abstract}
		
		\begin{keyword}
			\KWD \\Ultrasound image synthesis \\ Generative Adversarial Networks \\ COVID-19 \\ Hip joint \\ Ovary and follicle
		\end{keyword}
		
	\end{frontmatter}
	
	
	\section{Introduction} 
	Ultrasound (US) imaging is prevalent in routine clinical examinations because of its relatively low cost, real-time imaging capability and avoidance of radiation exposure \citep{kutter2009visualization, alessandrini2015pipeline}. During US diagnosis, sonographers first manually operate an imaging equipment to produce images required for diagnosis, and then review and analyze the images to find abnormalities \citep{doi2007computer}. This process relies heavily on sonographers' knowledge and experience. It usually takes a long time for novices to acquire operating and diagnostic skills. This is even truer when diagnosing rare diseases, due to the lack of training on real data \citep{mattausch2017comparison}.
	
	In recent years, we have witnessed considerable progress in computational medical image analysis for the detection, diagnosis, and treatment of diseases \citep{cheng2020computational}. Compared with medical image interpretation by human experts, automated analysis is more efficient, objective, and does not suffer from inter-observer variations \citep{cheng2016computer}. In the stream of applying machine learning, especially deep learning, to data analysis, large-scale datasets and annotations lie at the heart of its success to accomplish target tasks. For example, the ImageNet database, designed for visual object recognition, contains more than one million annotated images. However, in medical applications, usually only a very limited number of images are available, and annotations require expert knowledge about the data and task. Therefore, the lack of large-scale datasets and annotations remains a major obstacle hindering the successful application of deep learning algorithms to medical images \citep{liu2019deep, gao2019convolutional}.
	
	Researchers have been trying to circumvent this obstacle via data augmentation. The most common method is affine transformation, including translation, rotation, and scaling. This technique simply modifies original images to expand the dataset for model training. Although the sample size can be remarkably increased in this way, only little additional information is introduced into the dataset, due to the small content changes (e.g. rotating an image by an angle) \citep{frid2018gan, salehi2020generative}. In this regard, there is an urgent need for a new data augmentation method that can enrich the dataset with more variability, so that the model trained on a small dataset can also generalize well on unseen data \citep{yi2019generative}.
	
	Image synthesis is a new and more sophisticated augmentation method. It can be classified into physics-based and learning-based methods. Well-known US simulation packages such as Field II \citep{jensen1997field, jensen2004simulation} and k-Wave \citep{treeby2010k} can be used to simulate B-mode ultrasound images, though they are not only designed for image synthesis. k-Wave is designed for the time-domain simulation of propagating acoustic waves and can account for both linear and nonlinear wave propagation, while Field II is a linear ultrasound simulation tool. \citet{simulation2004} proposed a physical model for simulating intravascular US (IVUS) images. \citet{burger2012real} built deformable mesh models from CT volumes to fulfill real-time simulation by GPU. Cardiac US sequences were simulated based on an electromechanical model \citep{cardiacgeneration2012} and a warping strategy \citep{zhou2017framework}. Although these methods follow US imaging principles, their computational complexity is often high due to the modeling of wave propagation process \citep{freehand2017}. They are very time-consuming, especially for generating high-resolution images. Moreover, their performance may be affected by the quality of pre-built models, which are often essential and difficult to construct.
	
	In the past few years, deep learning based synthetic methods have gained more and more interest. Among them, the generative adversarial network (GAN) \citep{GAN} is the most promising approach. \citet{fujioka2019breast} employed a deep convolutional GAN (DCGAN) \citep{radford2015unsupervised} to synthesize breast US images without additional constrains. \citet{freehand2017} first proposed a novel spatially-conditioned GAN based on conditional GANs (cGANs) \citep{cGAN} to synthesize US images from fetal phantoms. The proposed architecture can improve the training stability by taking pixel coordinates as conditioning input. \citet{simulating2018} introduced a multi-stage method including two different cGANs to transform tissue maps into synthetic IVUS images. Although the cascaded cGANs are hard to train, this system is the first one to use tissue labels as conditioning input to enhance the training stability.
	
	Although cGAN \citep{cGAN, pix2pix} is effective and enables the user-controlled image generation, the synthesized images often have low resolution and checkerboard artifacts. To make the structural details of generated images more realistic, auxiliary guidance information, such as the sketch and edge of the background, was introduced \citep{shin2018abnormal, zhang2019skrgan}. However, it is still challenging to synthesize high-resolution images. Due to the more details in high-resolution images, the discriminator can easily recognize the differences between generated and real images, which may lead to the vanishing gradient problem and make the training difficult. Additionally, training such model is memory-intensive, which limits using a large batch size to improve training stability.
	
	\begin{figure*}[!t]
	\centering
	\includegraphics[width=\textwidth]{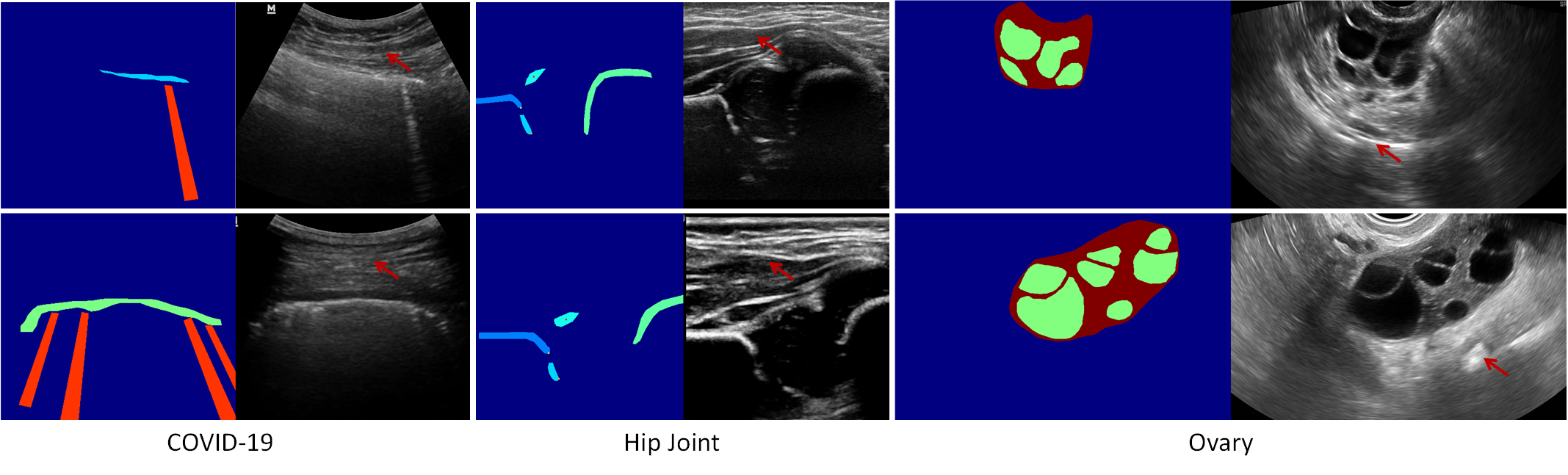}
	\caption{Examples of the US images and the corresponding label maps for three different datasets. On the left are the annotated label maps in pseudo colors, and on the right are the real US images. The red arrows indicate the non-target background regions in US images, which are difficult to synthesize realistically due to the lack of background information in the label maps.}
	\label{fig_datasets}
	\end{figure*}
	
	To address above issues, we devise a novel sketch guided and progressive growing GAN (spGAN) to synthesize US images. The main contributions of our work include:
	
	\begin{enumerate}
		
		\item
		To the best of our knowledge, this is the first work that can synthesize realistic B-mode US images with high-resolution and customized texture editing features. A software tool and a video demo of our method are available at GitHub
		
		(\href{https://github.com/Carmenliang/UI_synthesis}{https://github.com/Carmenliang/UI\_synthesis}). 
		\item
		To enhance the fidelity of synthesized structure details, we propose to introduce auxiliary sketch guidance into a cGAN. Specifically, we superpose the edge sketch onto the object mask and use the composite mask as the network input. Customized editing of the edge sketch and object mask makes our method quite flexible in generating different US images for training new sonographers and augmenting data in deep learning models.
		\item
		To generate high-resolution US images, we adopt a progressive training strategy \citep{PGGAN} to gradually generate high-resolution images from low-resolution images. In addition, a feature loss (FL) is proposed to minimize the difference of high-level features between the generated and real images, which further improves the quality of generated images.
		\item
		The proposed US image synthesis method is quite universal and can also be generalized to the US images of other anatomical structures besides the three ones tested in our study (lung, hip joint, and ovary).
		\item
		Extensive experiments on three large US image datasets are conducted to validate the efficacy of our method, including ablation studies, customized texture editing, user studies, and segmentation comparison between real and synthesized images.
	\end{enumerate}
	
	Some preliminary results of this study have been published in the ISBI 2020 conference \citep{liang2020synthesis}. In this paper, we make substantial extensions in the following two aspects. 1) A new regularization term is introduced into the loss function to make the generated images and real images alike in terms of high-level features, which can successfully remove the artifacts present in our previous study and other advanced GAN-based synthesis methods. 2) Besides the ovary dataset used in our previous study, we collect two additional large datasets of US images (COVID-19 and infant hip joint) to further validate our method. 3) We add segmentation experiments to demonstrate the efficacy of our method as a data augmentation approach. Compared with traditional data augmentation such as image translation and rotation, our GAN-based augmentation method can provide greater variability with editable operations and therefore has great potential to improve performance. 4) We add extensive ablation studies to verify the effectiveness of each component of our method. 5) We investigate the effect of three key parameters on the performance of our method. 6) We release our US image synthesis tool to a public repository (\href{https://github.com/Carmenliang/UI_synthesis}{https://github.com/Carmenliang/UI\_synthesis}), which can be readily used by other researchers.
	

	\section{Materials and methods}
	\label{Method}
	\subsection{Datasets}
	\label{Datasets}
	This study was approved by local institutional review boards. A robust US simulation framework is expected to be able to synthesize photo-realistic images with various characteristics, such as different structural shapes, positions, and echo patterns. Hence, three representative datasets of B-mode US images were collected and used in our study: lung US for diagnosis of COVID-19 (COVID-19), infant hip joint US (hip joint), and ovary and follicle US (ovary). Each of the three datasets has its own special characteristics. For the COVID-19 dataset, observing the specific echo patterns is important for diagnosis. For the hip joint dataset, attention is focused on the relative position of different anatomical structures. For the ovary dataset, doctors analyze the size of the ovary and the size and number of follicles.
	
	All US images had corresponding segmentation maps annotated by experienced doctors. Example images and the corresponding segmentation maps are shown in Fig.~\ref{fig_datasets}. The details of each dataset are described as follows:
	
	\textbf{COVID-19.}
	This dataset contained 6054 images totally, in which 4849 images were used as the training set and the remaining 1205 images as the test set. These images had different resolutions, with the height ranging from 179 to 799 pixels and width ranging from 109 to 1104 pixels. All images were resized to 256$\times$256 pixels and 512$\times$512 pixels for training low-resolution and high-resolution synthesis models. The annotated artifacts in lung US images included pleura line, A-line, B-line, and consolidation. The COVID-19 dataset was collected from multiple centers in Wuhan, including Cancer Center of Union Hospital, West of Union Hospital, Jianghan Cabin Hospital, Jingkai Cabin Hospital, and Leishenshan Hospital. Various ultrasound machines were used, including Mindray M7, M8, M9 and GE Logiq E9, Logiq E Portable Ultrasound Machine.
	
	\textbf{Hip Joint.}
	This dataset contained 1231 images totally, in which 992 images were used as the training set and the remaining 239 images as the test set. To remove the characters in original US images, we cropped them to 512$\times$512 pixels and then resized them to 256$\times$256 pixels. Both cropped and resized images are used for training the backbone structure of GAN. Four structures were annotated in the segmentation maps, including ilium, lower limb, labrum, and co-junction. The hip joint dataset was collected from Guangdong Women and Children Hospital with two different machines (Hitachi HI-Vision Preirus and Philips iU22). The frequencies of Hitachi’s transducer are between 5-13 MHz, while the frequencies of the other one are between 3-9 MHz.
	
	\textbf{Ovary.}
	This dataset contained 3261 ovarian images totally, in which 2848 images were used as the training set and the remaining 413 as the test set. The image size was non-uniform with the height ranging from 380 to 530 pixels and width ranging from 610 to 860 pixels. All images were resized to 256$\times$256 pixels and 512$\times$512 pixels for training. Ovary and follicles were annotated in the segmentation maps. The ovary dataset was collected from The Third Affiliated Hospital of Guangzhou Medical University with two different machines (Mindray Resona 7S and GE Voluson 6S). The frequencies of Mindray’s transvaginal transducer are between 3-9 MHz, while the frequencies of the other one are between 4-10 MHz.
	
	\begin{figure*}[!t]
		\centering
		\includegraphics[width=0.8\textwidth]{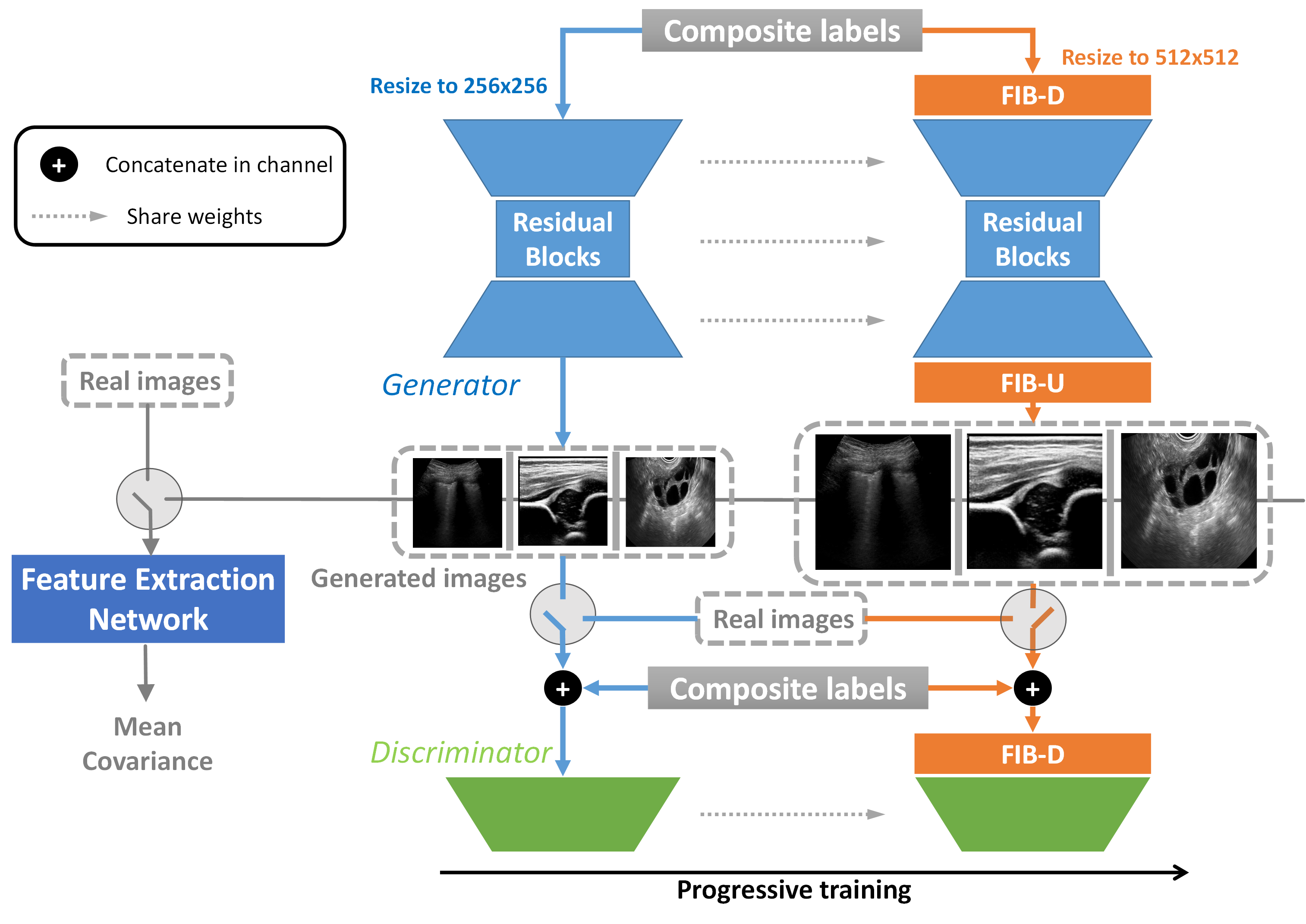}
		\caption{Overview of our proposed US image synthesis framework. The composite labels contain original annotated structures as well as auxiliary sketches. For the generator, the input is the composite labels and output is the generated US images. For the discriminator, the input is the US images and the corresponding composite labels. In the progressive training scheme, the left backbone structure is adopted as a pre-trained model for low-resolution image (256$\times$256) synthesis, and then fade-in blocks (FIB-D, FIB-U) are added for synthesizing realistic high-resolution images (512$\times$512). Moreover, the additional feature extraction network is employed for calculating the mean and covariance of the generated and real images, and then they are used for constructing the feature loss aiming at improving synthesis quality further.}
		\label{fig_framework}
	\end{figure*}

	\subsection{Overview of our proposed method}
	\label{Overview}
	To synthesize high-fidelity and high-resolution US images from simple segmentation maps, we proposed a GAN-based image synthesis framework, as illustrated in Fig.~\ref{fig_framework}. Specifically, we first added fine-grained edge sketch to original label maps, which resulted in the composite label maps that can help the generator create images with realistic texture (Section~\ref{Guidance}). Next, the designed backbone GAN structure was used for a warm-up training of low-resolution images of size 256$\times$256 (Section~\ref{Backbone}). Then, a progressive growing scheme is introduced for synthesizing high-resolution images (Section~\ref{Progressive}). To enable a smooth transition between layers, FIBs were incorporated into the backbone structure (Section~\ref{FIB}). Finally, a feature loss was employed to further improve the texture fidelity of synthesized images (Section~\ref{Feature_loss}). The above key components of our image synthesis framework are described in detail in the following sections.

	\subsection{Auxiliary guidance}
	\label{Guidance}
	
	\begin{figure*}[!t]
		\centering
		\includegraphics[width=\textwidth]{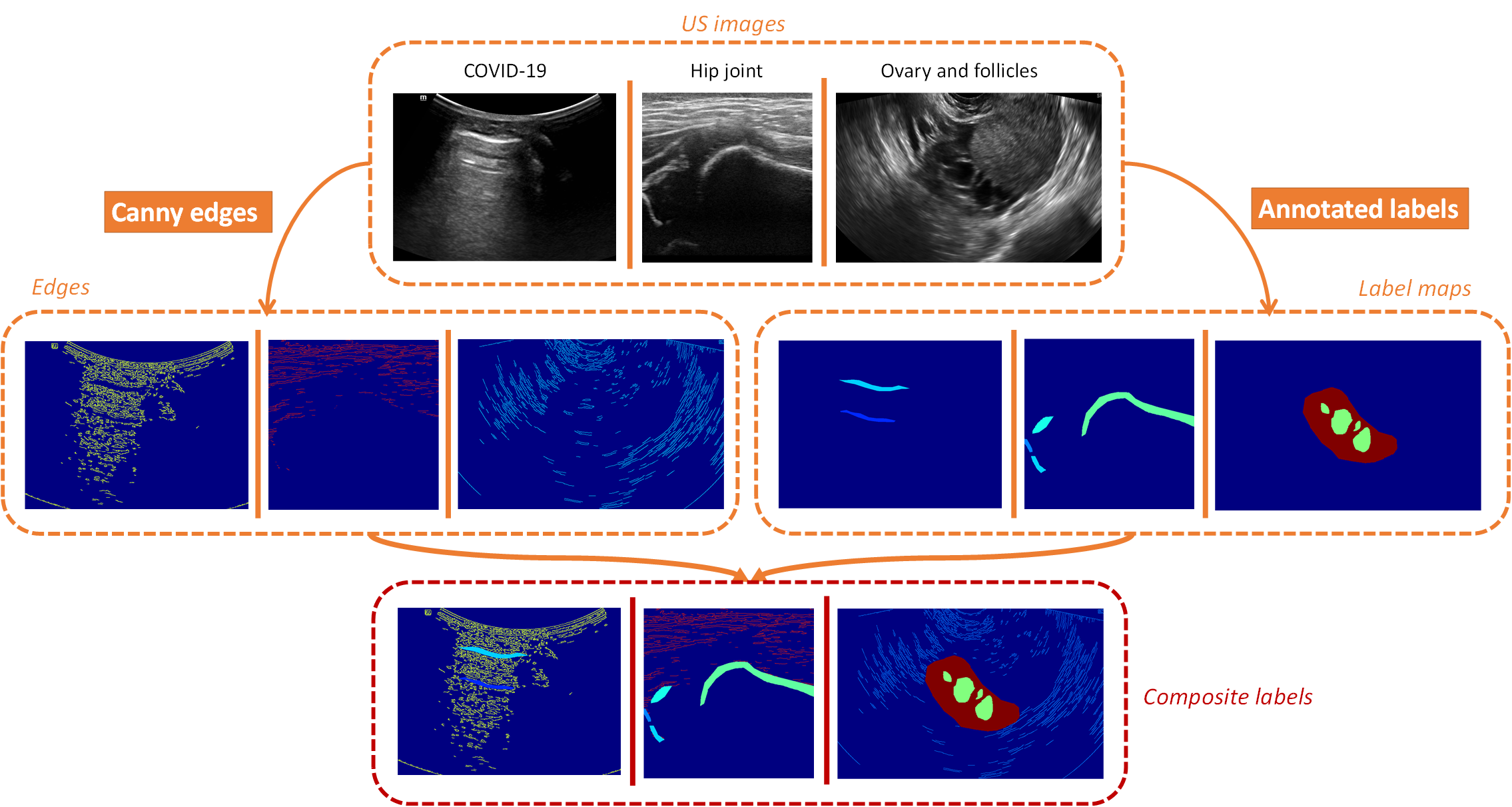}
		\caption{Generation of composite labels by adding the edge sketches onto the original label maps.}
		\label{fig_composite}
	\end{figure*}
	
	There are two common image-to-image translation tasks in the field of medical image analysis: translation between different imaging modalities (such as MRI T1-to-T2 \citep{liu2019susan, dar2019image}, CT-to-MRI \citep{jiang2018tumor}, MRI-to-CT \citep{nie2017medical, zhao2018craniomaxillofacial}, CT-to-PET \citep{ben2017virtual}, PET-to-MRI \citep{choi2018generation}) and transformation from segmentation maps to medical images. Image synthesis from segmentation maps can generate different images by simply modifying the content in the maps, which is a desirable feature for data augmentation. Since segmentation maps only contain the shape of target structures and lack background details, the transformation from segmentation maps to medical images is generally more difficult than translation between different imaging modalities. As for US images, synthesizing from segmentation maps is even harder due to a large amount of noise in US images. Motivated by previous work \citep{shin2018abnormal, zhang2019skrgan}, we used the edge information of the background texture as auxiliary sketch guidance to achieve high-fidelity synthesis and customized editing of US images. 
	
	Specifically, the Canny algorithm \citep{canny} was applied to real images to extract binary edge sketch because it is robust against noise. We then updated the original segmentation map of the target object $\textbf{O}$ by superposing the edge sketch $\textbf{S}$ onto it, resulting in the composite label $\tilde{\textbf{O}}$, which is defined as:
	
	\begin{align}
	\label{eq:composite}
	\tilde{O} = M\otimes O + (1-M)\otimes S
	\end{align}
	where \textbf{M} ($\textbf{M} \in \{0, 1\}$) denotes the binary map indicating the area for annotated structures. \textbf{$\otimes$} refers to the operation of element-wise multiplication.
	Through the above operation, the auxiliary sketch $\textbf{S}$ is superposed onto the original mask $\textbf{O}$ without affecting the area of the target objects, as shown in Fig.~\ref{fig_composite}. With the additional auxiliary sketch of background provided for GAN to learn, our method can generate images with realistic background texture.

	\subsection{Backbone structure of GAN}
	\label{Backbone}
	The backbone structure of GAN synthesizing images of size 256$\times$256, as shown in the middle of Fig.~\ref{fig_framework}, is a basic structure for subsequent high-resolution image synthesis.
	The composite label generated in the previous section was used as the input of both the generator and discriminator of the GAN structure.
	
	For the generator, the conditional input was the composite labels and the output was the synthesized US images. We utilized the encoder-decoder architecture with $n$ residual blocks in between. The architecture of the encoder and decoder included three down-sampling blocks and three up-sampling blocks, respectively. Each up-sampling or down-sampling block was comprised of one convolution or deconvolution layer with a stride of 2, one instance normalization (IN) \citep{ulyanov2016instance} layer, and one ReLU activation function.
	
	The number of residual blocks controls the ability of feature extraction in the model, and therefore can be different for specific datasets. Each residual block contained two convolution layers with a stride of 1. Except for the first and the last convolution layers with a kernel size of 1$\times$1 for channel size adaptation, all other convolution and deconvolution layers used a kernel size of 3$\times$3. 
	
	For the discriminator, we used the PatchGAN \citep{pix2pix}, whose input was the concatenation of the composite labels and generated/real images. The PatchGAN can be designed to different output sizes. Each unit of the output reflects the possibility of an image patch being real, which is used to calculate adversarial loss when training.
	Because PatchGAN places more restrictions on local regions and has more high-frequency information to feed back to the generator, it often performs better than the image-based discriminator.
	Therefore, we used the PatchGAN structure as the discriminator, which consists of five convolution layers. 
	
	The objective function of generator $L_{G}$ was composed of a conditional adversarial loss $L_{GAN_G}$ and a L1-loss $L_{L1}$ for low-frequency restriction, which are formulated as follows:
	
	\begin{align}
	\label{eq:adversarial_loss_g}
	L_{GAN_G}(G) &= E_{x,G(x)}[log(1-D(x,G(x)))] \\
	\label{eq:L1_loss}
	L_{L1}(G) &= \lVert y - G(x) \rVert _1\\
	\label{eq:g_total_loss}
	L_{G}(G) &= L_{GAN_G} + \lambda_{1} L_{L1}
	\end{align}
	where $G, D$ represent the generator and discriminator respectively, $x$ denotes the conditional composite labels, $y$ denotes the ground truth US images, $G(x)$ denotes the synthesized images from input $x$. The hyperparameter $\lambda_{1}$ was set to 1. The objective function of discriminator $L_{D}$ is calculated as:
	
	\begin{align}
	\label{eq:d_loss}
	L_{D}(D) &= E_{x,y}[log D(x,y)] \notag\\
	&+E_{x,G(x)}[log(1 - D(x,G(x)))]
	\end{align}
	
	The objective function of the discriminator was maximized while the objective function of the generator was minimized during the adversarial training process. The discriminator training alternated with generator training in each epoch. This alternating process was repeated until the generated images were sufficiently realistic.
	
	\subsection{Progressive growing scheme}
	\label{Progressive}
	Compared to low-resolution images (256$\times$256), high-resolution images (512$\times$512) have more fine structures. The backbone architecture alone as described in Section~\ref{Backbone} is not capable of extracting enough information for generating high-resolution, realistic images. In order to synthesize high-resolution US images with high fidelity, we adopted the progressive growing scheme \citep{PGGAN} to decompose the task as incremental learning ones. This scheme enables us to use only one generator and one discriminator with fast and smooth learning for high-resolution, realistic synthesis. Specifically, we started from an easier task that synthesizes low-resolution images in several warm-up epochs with the backbone structure, and then, the weights of the backbone structure were shared with the generator and discriminator for high-resolution US image synthesis.
	
	The entire training process can be divided into four phases. In phase 1, the backbone architecture of generator and discriminator (Section~\ref{Backbone}) was applied for low-resolution (256$\times$256) US images synthesis. After several warm-up epochs until convergence, this pre-trained architecture enabled good quality synthesis of low-resolution US images. In phases 2 and 3, we trained the discriminator and generator for high-resolution image synthesis, sequentially and respectively. In this process, new layers were added to the networks, and we faded them in smoothly with the FIBs to avoids sudden shocks to the already well-trained layers (see Fig.~\ref{fig_framework}). The discriminator was trained earlier than the generator to replenish the gradient information and force the generator to learn to synthesize higher resolution images. Finally, in phase 4, we trained the discriminator and generator together (the rightmost part of Fig.~\ref{fig_framework}) for several more epochs to further enhance performance.
	
	\subsection{Fade-in Blocks}
	\label{FIB}
	
	\begin{figure}[!t]
		\centering
		\includegraphics[width=0.5\textwidth]{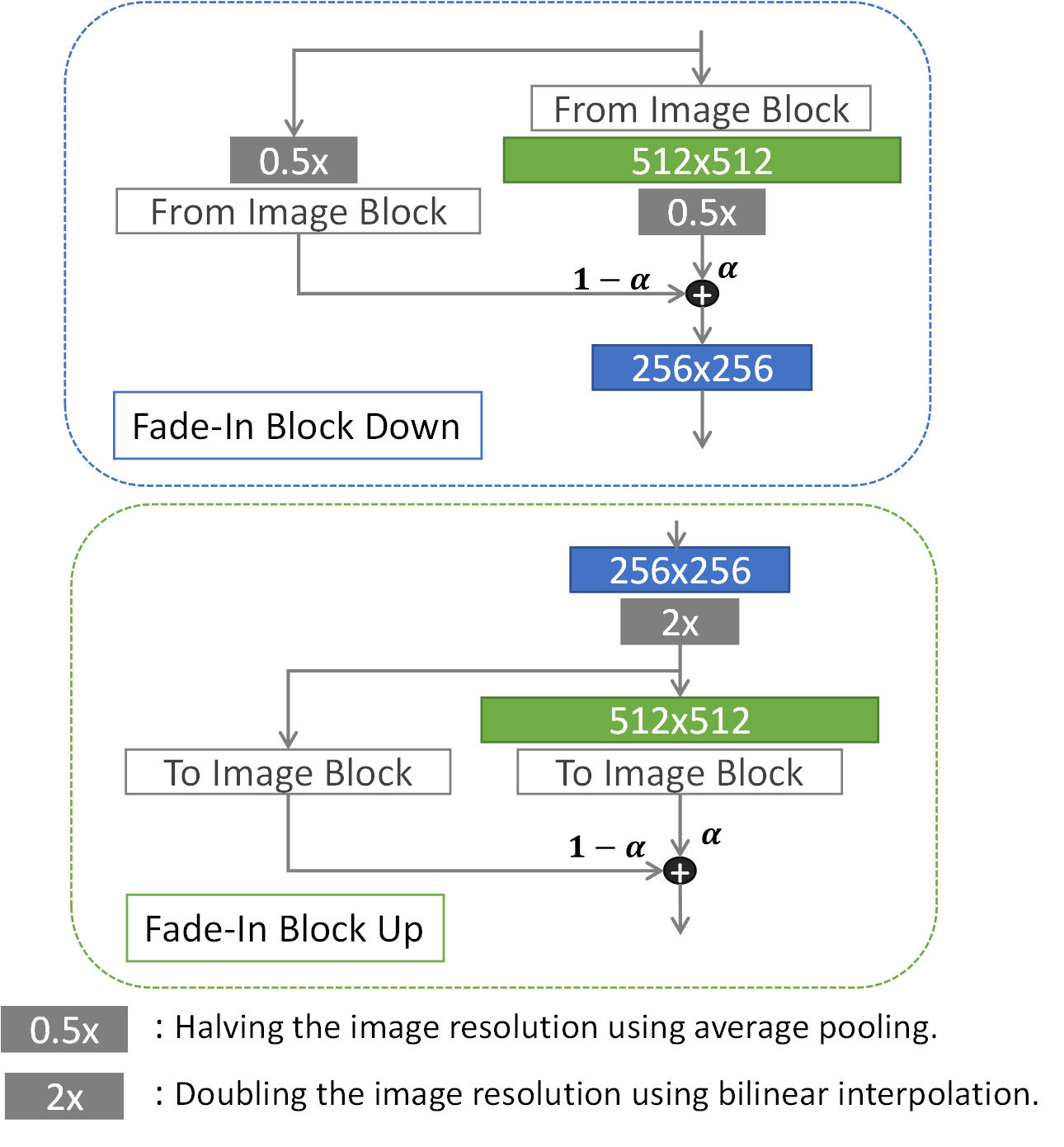}
		\caption{Diagram of the fade-in block down (top) and fade-in block up (bottom). These two blocks are used in the progressive growing scheme, with $\alpha$ increasing during transition phases. The from image block represents a layer projecting image channels to feature vectors using 1$\times$1 convolution, and the to image block functions the opposite way. The block 512$\times$512 contains two 3$\times$3 convolution layers, and the block 256$\times$256 represents the original structures of backbone adjacent to newly added structures.}
		\label{fig_FIB}
	\end{figure}
	
	To avoid the sudden shock during training when new layers are added, FIBs were adopted for a smooth transition in both generator and discriminator. Fig.~\ref{fig_FIB} illustrates the structure of FIB. Following the design of residual neural network, FIB also uses a skip connection (Fig.~\ref{fig_FIB}). The side branch skips over the convolutional layers in the main branch and then merges into the main branch through a weighted sum. The weight $\alpha$ ($\alpha \in [0, 1]$) controls the balance between the main and side branches. $\alpha = 0$ indicates that only the side branch determines the output of FIB, while $\alpha = 1$ means that the output only depends on the main branch. The use of FIB not only smooths the transition between different resolutions but also makes weights sharing possible and training more efficient. 
	
	Two kinds of FIB were designed for the purpose of down-sampling and up-sampling and denoted as FIB-Down (FIB-D) and FIB-Up (FIB-U), respectively. The difference between FIB-D and FIB-U is the direction of resizing. For FIB-D, it halves the resolution using average pooling, and for FIB-U, it applies bilinear interpolation for doubling the resolution. FIB-D was used in both generator and discriminator while FIB-U was only used in the generator (Fig.~\ref{fig_framework}). 
	
	Fig.~\ref{fig_FIB} shows the two-branch structure in FIB. The lightweight side branch helps the pre-trained network adapt easily. The more complex architecture of the main branch has stronger feature extraction capabilities. Combining these advantages, the $\alpha$ is introduced to guide the side branch to gradually switch to the main one. Specifically, it is increased from 0 to $\alpha _{max}$ with a fixed step. It is noted that updating the $\alpha$ in both discriminator and generator simultaneously may decrease the stability and thus we increased $\alpha$ alternately when training the two modules. Besides, we also noticed that $\alpha _{max}$ had an impact on the performance of the generator. A lower $\alpha _{max}$ may limit feature extraction capabilities, while a larger one tends to cause a sudden increasing loss. In comparison, the loss in the discriminator varies smoothly even with a larger $\alpha _{max}$. For these reasons, we set the $\alpha _{max}$ to 0.5 and 1.0 in generator and discriminator, respectively.
	
	\subsection{Feature loss}
	\label{Feature_loss}
	By employing the sketch guidance in segmentation maps, progressive training scheme and FIBs, we achieved editable, high-resolution synthesis with undistorted texture.
	However, there were some small flaws in generated high-resolution images, which are also commonly seen in other GAN-based synthesis methods. 
	The generated images appeared to be a little blurry compared with ground truth images and had some salt and pepper noise especially in the regions without annotated labels and sketches.
	The flaws might be caused by the pixel-wise L1-loss when training the generator, since the loss only restricts pixel-wise difference and neglects the spatial similarity between pixels \citep{blau2018perception}. However, removing the L1-loss term from the objective function led to poor performance due to the lack of low-frequency restriction.
	
	Therefore, we proposed a feature loss to add extra restrictions for better texture synthesis.
	Inspired by the perceptual loss \citep{johnson2016perceptual} originally used in style transfer and super-resolution tasks, we calculated the high-level features of the real and generated images and tried to minimize the mean and covariance between them. To extract high-level features, images were fed to a ResNet-50 \citep{he2016deep} model which was trained with 1500K prenatal US images for standard plane detection, and the output of the layer conv4 of the model was used as the high-level features. We referred to the feature extraction layers in the ResNet-50 as feature extraction network (FEN). The feature loss $L_{F}$ is given by:
	
	\begin{align}
	\label{eq:LF_loss}
	L_{F}(G) &= \lVert Mean(FEN(x)) - Mean(FEN(G(x))) \rVert _1 \notag\\
	&+\lVert Var(FEN(x)) - Var(FEN(G(x))) \rVert _1
	\end{align}
	
	The updated objective function of the generator is given by:
	
	\begin{align}
	\label{eq:update_g_loss}
	L_{G}(G) &= L_{GAN_G} + \lambda_{1} L_{L1} + \lambda_{2} L_{F}
	\end{align}
	The hyperparameter $\lambda_{2}$ was set to different values for different datasets, which will be discussed in Section~\ref{Result}.
	
	\begin{figure*}[h]
		\centering
		\includegraphics[width=0.8\textwidth]{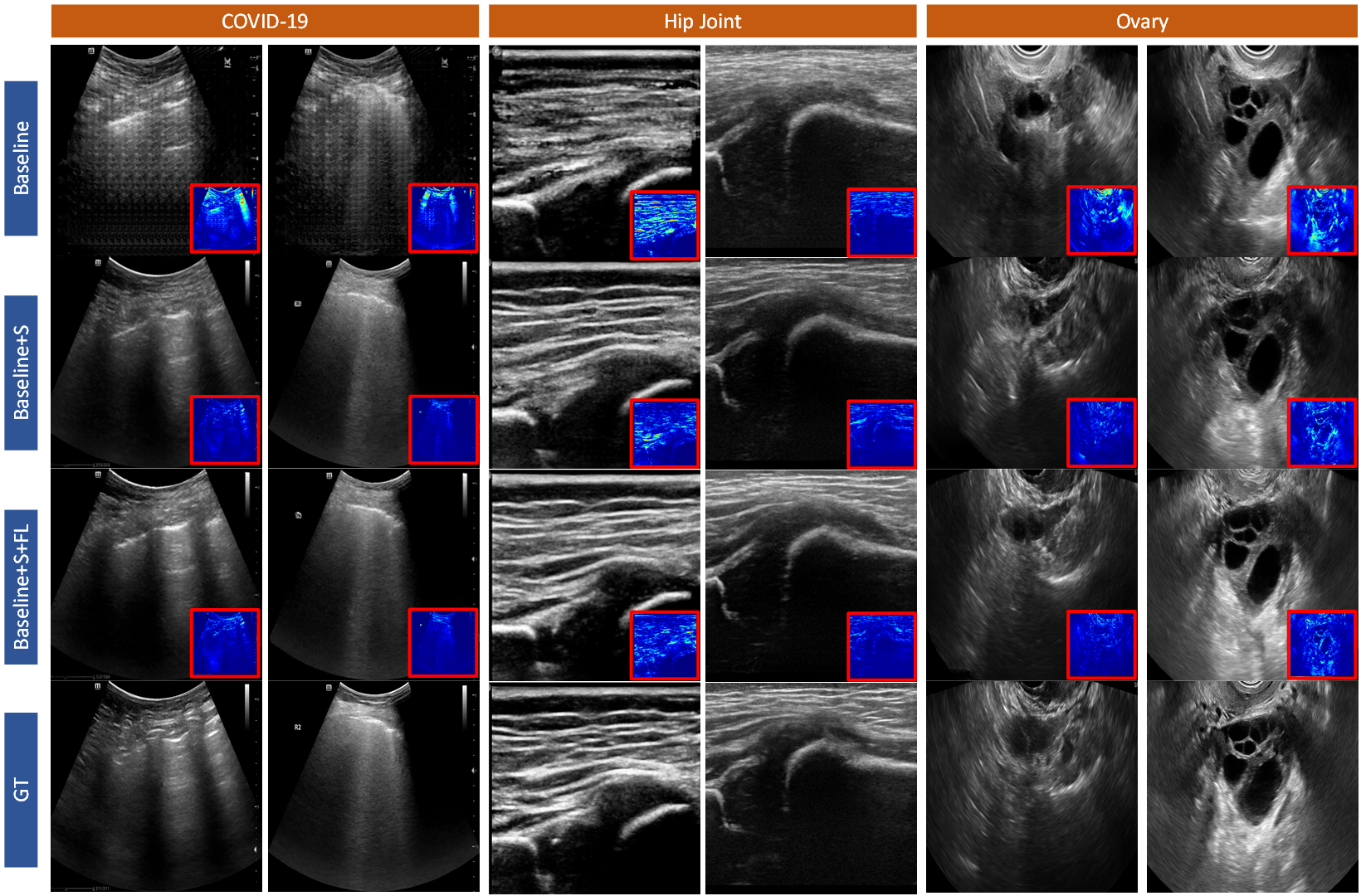}
		\caption{Examples of low-resolution (256$\times$256) image synthesis for three datasets. The first three rows are results for three different methods, and the last row presents the GT images. In the bottom right corner of each synthesized image is the difference heatmap. S, sketch; FL, feature loss.}
		\label{fig_256}
	\end{figure*}

	\section{Experiments and results}
	\label{Result}
	
	\subsection{Implementation details}
	\label{Details}
	Our experiments were implemented using Pytorch with a single NVIDIA GeForce RTX 2080 Ti GPU. The proposed spGAN can be split into four phases, as described in Section~\ref{Progressive}. In phase 1, we first trained the backbone structure of GAN for low-resolution image synthesis. We used a batch size of 4 with Adam optimizer. The learning rates of generator and discriminator were set to 0.001 and 0.0001, respectively. The weight of the feature loss term in the objective function of the generator $\lambda_{2}$, the number of residual blocks in the generator $n$, and the output size of PatchGAN $s$ were 10, 15, and 30$\times$30 for the COVID-19 dataset, 10, 15, and 120$\times$120 for the hip joint dataset, and 5, 10, and 30$\times$30 for the ovary dataset. Note the above settings were determined by grid search and used in all experiments unless specified otherwise. For the COVID-19 and ovary datasets, the models in phase 2 and 3 were trained for 50 epochs, with $\alpha$ increasing by a step of 1/50. For the hip joint dataset, the models in phase 2 and 3 were trained for 100 epochs, with $\alpha$ increasing by a step of 1/100. In phase 4, the full model was trained for 200 epochs for the COVID-19 and ovary datasets, and for 400 epochs for the hip joint dataset.
	
	We evaluated the performance of our proposed method on three US datasets, including COVID-19, hip joint, and ovary images. Both qualitative and quantitative results were presented in the experiments. For qualitative results, we compared the real US images and synthesized images and showed the heatmap of pixel-level differences between them. For quantitative evaluation, four numerical metrics were adopted, including freshet inception distance (FID) \citep{FID}, kernel inception distance (KID) \citep{KID}, multi-scale structural similarity (MS-SSIM) \citep{MS-SSIM}, and learned perceptual image patch similarity (LPIPS) \citep{LPIPS}. Both FID and KID measure the distance between images at the feature level. The difference is that KID estimates are unbiased. The lower value of the FID and KID indicates better performance. MS-SSIM measures the similarity of the paired images, ranging from 0 to 1, where larger value means better performance. For LPIPS, we used a pre-trained ResNet-50 \citep{he2016deep} network to calculate the perceptual differences in multiple layers, with smaller differences meaning better performance.

	\subsection{Low-resolution image synthesis}
	\label{Low-resolution}
	In this section, we presented the results of low-resolution (256$\times$256) image synthesis for different US datasets.
	To demonstrate the effectiveness of the sketch guidance and feature loss used in our synthesis framework, we compared the results of the following three methods: baseline, baseline+S, and baseline+S+FL. The baseline only used the backbone structure of GAN (Section~\ref{Backbone}), baseline+S incorporated the auxiliary sketch guidance (Section~\ref{Guidance}), and baseline+S+FL additionally incorporated the feature loss (Section~\ref{Feature_loss}).
	Due to different number of training samples and different texture complexity in the three datasets, the three methods were trained for different number of epochs: 150, 350, and 300 for the COVID-19, hip joint, and ovary datasets, respectively.
	
	The qualitative results are shown in Fig.~\ref{fig_256}, and the difference heatmap between the generated and ground truth (GT) images is shown on the bottom right corner of the generated images. The color in the heatmap from blue to red corresponds to the difference from small to large. As shown in Fig.~\ref{fig_256}, the addition of sketch guidance to baseline remarkably improved the quality of synthesized images. The checkerboard pattern in the generated COVID-19 images by the baseline method was successfully removed when the baseline+S method was used.
	Moreover, we saw further visual improvement when the feature loss was applied. Some scattered dark dots were observed in the generated COVID-19 images with the baseline+S method. However, they were removed successfully when the baseline+S+FL was used.
	
	More clearly, the quantitative results in Table~\ref{table_low_res} show a large improvement regarding all performance metrics for all datasets when the sketch guidance was added to the baseline.
	Addition of the feature loss to the baseline+S method further improved the performance for all datasets in terms of all metrics except the MS-SSIM and LPIPS for the hip joint dataset.
	
	\begin{table}[!h]
		\scriptsize
		\setlength{\abovecaptionskip}{0pt}
		\setlength{\belowcaptionskip}{10pt}
		\caption{Quantitative results for low-resolution (256$\times$256) image synthesis. S, sketch; FL, feature loss.}
		\label{table_low_res}
		\centering
		\begin{tabular}{lllllll}
			\toprule
			
			Dataset & Setting & FID $\downarrow$ & KID$\times$100 $\downarrow$ & MS-SSIM $\uparrow$ & LPIPS $\downarrow$\\
			
			\midrule
			\multirow{3}{*}{COVID-19} &
			Baseline & 196.98 & 40.14 & 0.3474 & 0.3272\\		
			& Baseline+S & 65.41 & 14.47 & 0.7428 & 0.2103 \\
			& Baseline+S+FL & \textbf{60.75} & \textbf{13.36} & \textbf{0.7450} & \textbf{0.1693} \\
			
			\midrule
			\multirow{3}{*}{Hip Joint} &
			Baseline & 98.51 & 6.70 & 0.4966 & 0.7139 \\
			& Baseline+S & 82.30 & 3.13 & \textbf{0.7151} & \textbf{0.4148} \\
			& Baseline+S+FL & \textbf{67.68} & \textbf{1.45} & 0.6853 & 0.4401 \\
			
			\midrule
			\multirow{3}{*}{Ovary} &
			Baseline & 67.80 & 8.45 & 0.2386 & 1.2319\\		   
			& Baseline+S & 63.15 & 5.57 & 0.4332 & 0.8901\\
			& Baseline+S+FL & \textbf{61.48} & \textbf{5.36} & \textbf{0.5035} & \textbf{0.8177} \\
			
			\bottomrule
			
		\end{tabular}
	\end{table}

	\subsection{High-resolution image synthesis}
	\label{Ablation}
	For synthesizing realistic, high-resolution (512$\times$512) US images, our proposed spGAN included three key components: the use of the auxiliary sketch guidance in segmentation maps, the pre-trained model for low-resolution image synthesis, and progressive growing of the $\alpha$ in the FIB.
	The proposed spGAN v2 included an additional key component, the feature loss. We conducted extensive ablation studies to evaluate these components. The proposed spGAN was compared with three variants, each with one of the three components removed as shown in Table~\ref{table_ablation}: spGAN-sketch, spGAN-pretrain, and spGAN-growing. Further, we compared spGAN and spGAN v2 to explore the effect of the feature loss. In addition, we also included as a comparison method the bilinear interpolation result of low-resolution images generated by baseline+S+FL (see Section~\ref{Low-resolution}), and denoted it as 256+interp.
	
	The quantitative results are presented in Table~\ref{table_ablation}. Compared with the interpolation results based on the generated low-resolution images, the proposed spGAN v2 achieved large performance gains regarding almost all evaluation metrics for all three datasets, except the MS-SSIM metric for the hip joint dataset. This means that the proposed high-resolution image synthesis framework can learn feature representations with more realistic details than simple image interpolation.
	
	By removing any one of the three components used in the spGAN, the performance was degraded in most cases, which confirms the effectiveness of these adopted components. Of the three components, the most important one is the auxiliary sketch guidance. Removing the sketch guidance led to a significant decrease in performance for all metrics and for all datasets. This indicates that the sketch guidance is essential for the model to learn convincing texture.
	
\begin{figure*}[!t]
		\centering
		\includegraphics[width=0.9\textwidth]{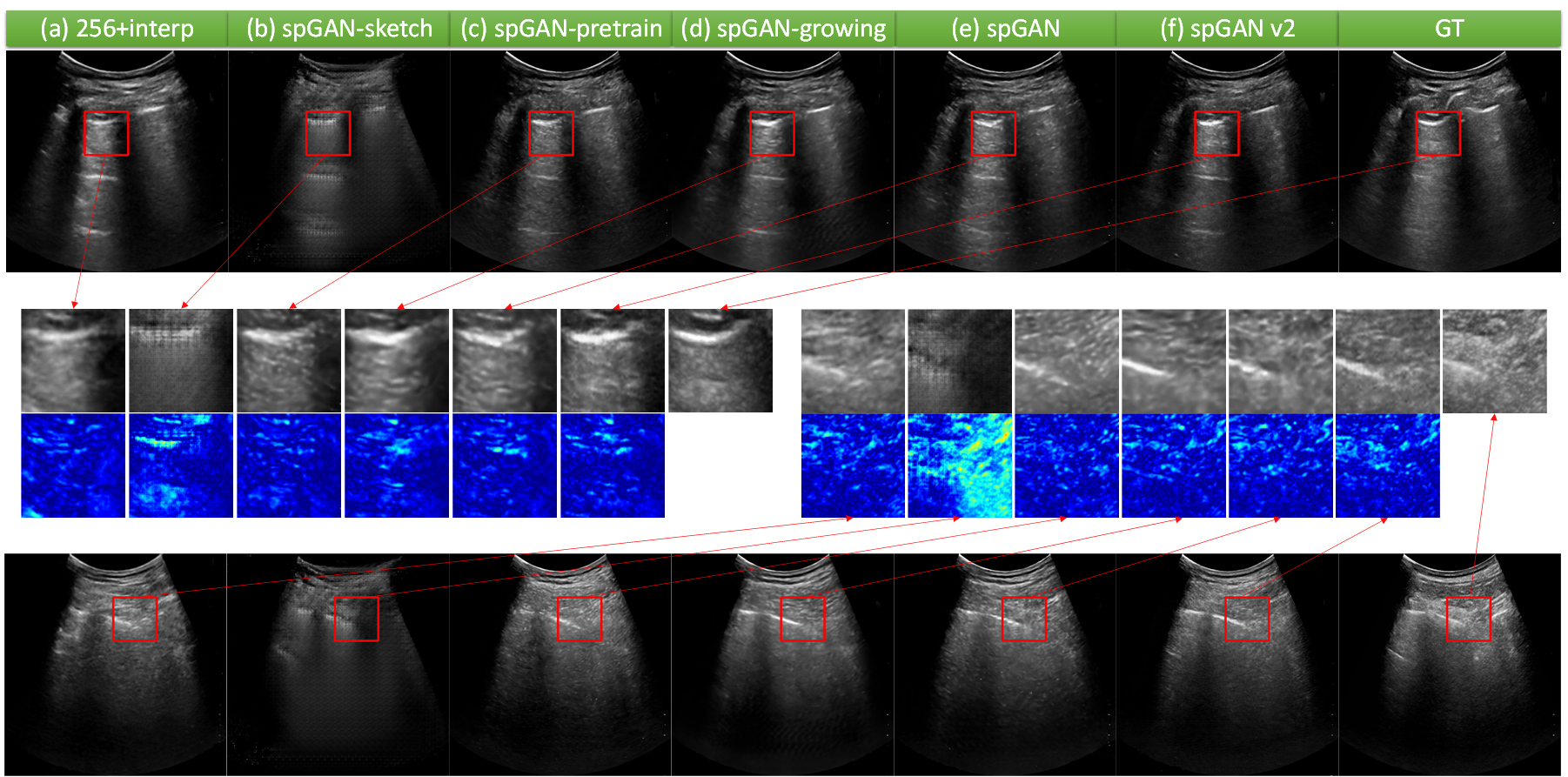}
		\caption{Examples of high-resolution (512$\times$512) image synthesis for the COVID-19 dataset using different methods. The middle two rows show the enlarged patches and the corresponding difference heatmap. Interp, interpolation.}
		\label{fig_512_covid}
	\end{figure*}

\begin{figure*}[!t]
	\centering
	\includegraphics[width=0.9\textwidth]{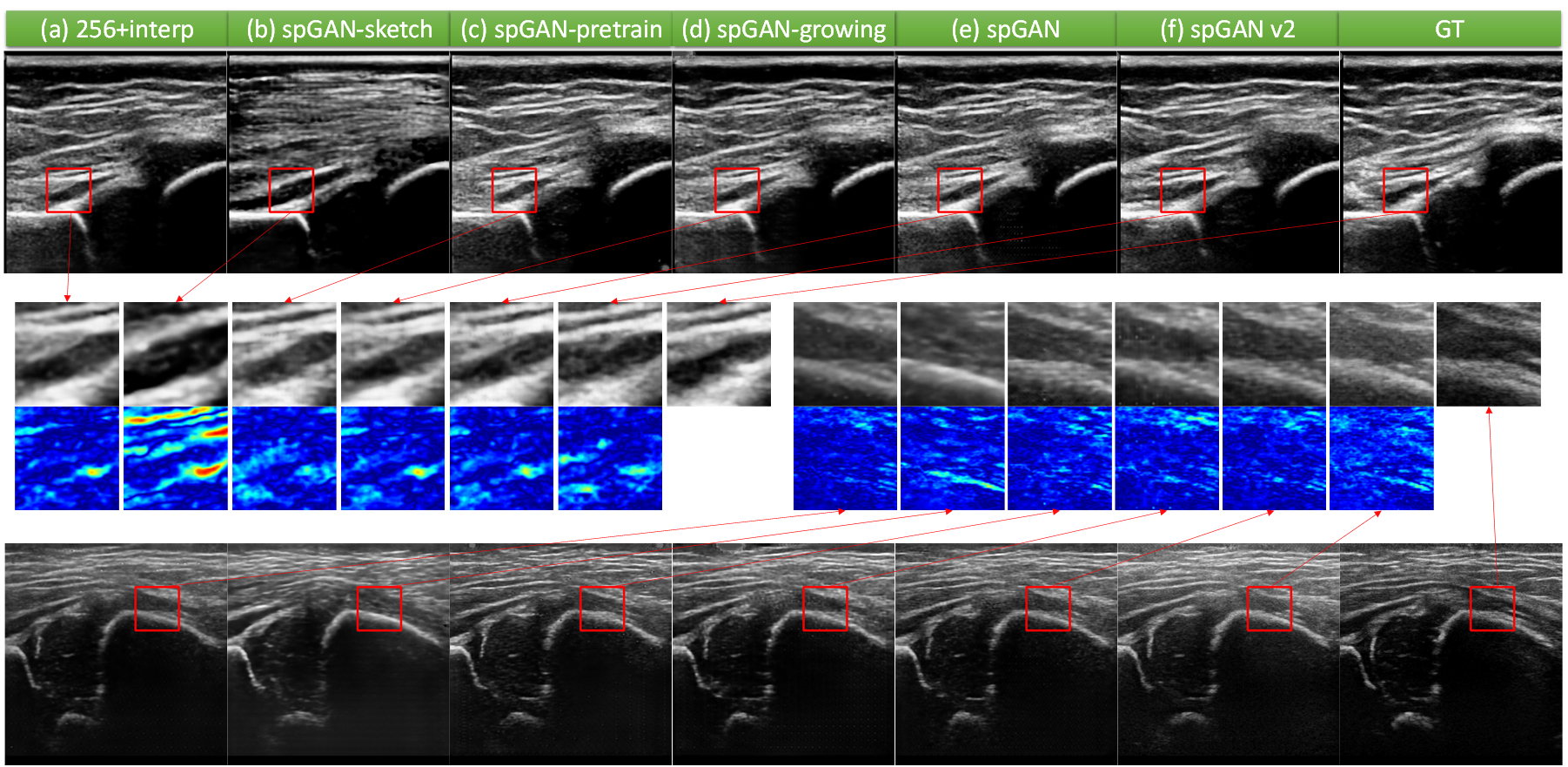}
	\caption{Examples of high-resolution (512$\times$512) image synthesis for the hip joint dataset using different methods. The middle two rows show the enlarged patches and the corresponding difference heatmap. Interp, interpolation.}
	\label{fig_512_hip}
	\end{figure*}

\begin{figure*}[!t]
	\centering
	\includegraphics[width=0.9\textwidth]{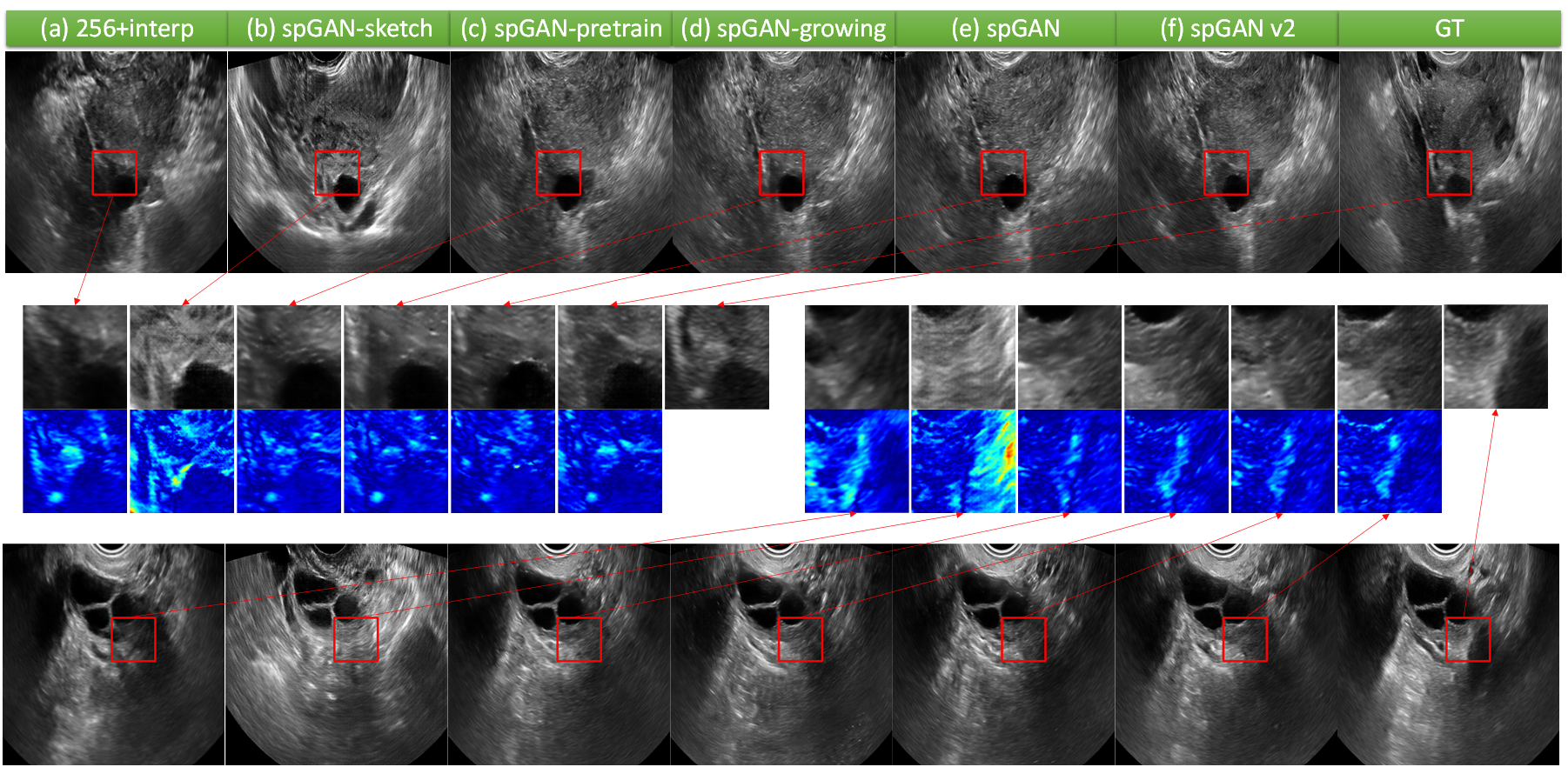}
	\caption{Examples of high-resolution (512$\times$512) image synthesis for the ovary dataset using different methods. The middle two rows show the enlarged patches and the corresponding difference heatmap. Interp, interpolation.}
	\label{fig_512_ovary}
	\end{figure*}

	Fig.~\ref{fig_512_covid}, \ref{fig_512_hip}, \ref{fig_512_ovary} show some examples of the generated images and difference heatmaps for the COVID-19, hip joint, and ovary images, respectively. Consistent with quantitative results, we observed that the proposed spGAN produced images that were more realistic compared with other baselines in most cases.
	\begin{table*}[!h]
		\scriptsize
		\setlength{\abovecaptionskip}{0pt}
		\setlength{\belowcaptionskip}{10pt}
		\caption{Quantitative results for high-resolution (512$\times$512) image synthesis. Interp, interpolation.}
		\label{table_ablation}
		\centering
		\begin{threeparttable}
			\begin{tabular}{l|llll|llll|llll}
				\hline
				
				\multirow{2}{*}{Setting} & \multicolumn{4}{c|}{COVID-19} & \multicolumn{4}{c|}{Hip Joint} &\multicolumn{4}{c}{Ovary}
				
				\\
				& FID$\downarrow$ & KID \newline$\times$100$\downarrow$ & MS-SSIM$\uparrow$ & LPIPS$\downarrow$ & FID$\downarrow$& KID\newline $\times$100$\downarrow$ & MS-SSIM$\uparrow$ & LPIPS$\downarrow$ & FID$\downarrow$& KID\newline $\times$100$\downarrow$ & MS-SSIM$\uparrow$ & LPIPS$\downarrow$\\
				\hline
				(a) 256+interp &  
				65.41 & 14.47& 0.7428 & 0.2103 &
				82.30 & 3.13 & 0.7151 & 0.4148 &
				63.15 & 5.57 & 0.4332 & 0.8901 \\
				\hline
				
				(b) spGAN-sketch & 
				156.89 & 34.96 & 0.4617& 0.2355 &
				107.89 & 7.87 & 0.5034 & 0.7188 &
				90.07 & 12.85 & 0.2254 & 1.2836 \\
				
				(c) spGAN-pretrain &  
				60.39 & 12.92 & 0.7501 & 0.1808 &
				82.75 & 3.16 & 0.7261 & \textbf{0.3519} &
				56.96 & 4.37 & 0.4966 & 0.7932 \\
				
				(d) spGAN-growing &
				59.45 & 12.81 & \textbf{0.7640}& 0.1820 &
				94.00 & 4.85 & 0.7090 & 0.4319 &
				63.13 & 5.35 & 0.4892 & 0.7776\\
				
				\hline
				(e) spGAN & 
				51.86 & 11.38 & 0.7590 & 0.1711 &
				
				77.04 & 2.26 & \textbf{0.7334} & 0.4011 &
				55.00 & 4.10 & 0.4879 & 0.7612 \\
				
				(f) spGAN v2 & 
				\textbf{36.36} &\textbf{8.04} & 0.7515 & \textbf{0.1682} &
				\textbf{57.81} & \textbf{0.20} & 0.7008 & 0.3947 &
				\textbf{47.11} & \textbf{2.67} & \textbf{0.4970} & \textbf{0.7950}\\
				\hline	
			\end{tabular}	
			
		\end{threeparttable}
	\end{table*}

	\subsection{User studies}
	\label{User_study}
\begin{figure*}[!t]
		\centering
		\includegraphics[width=0.9\textwidth]{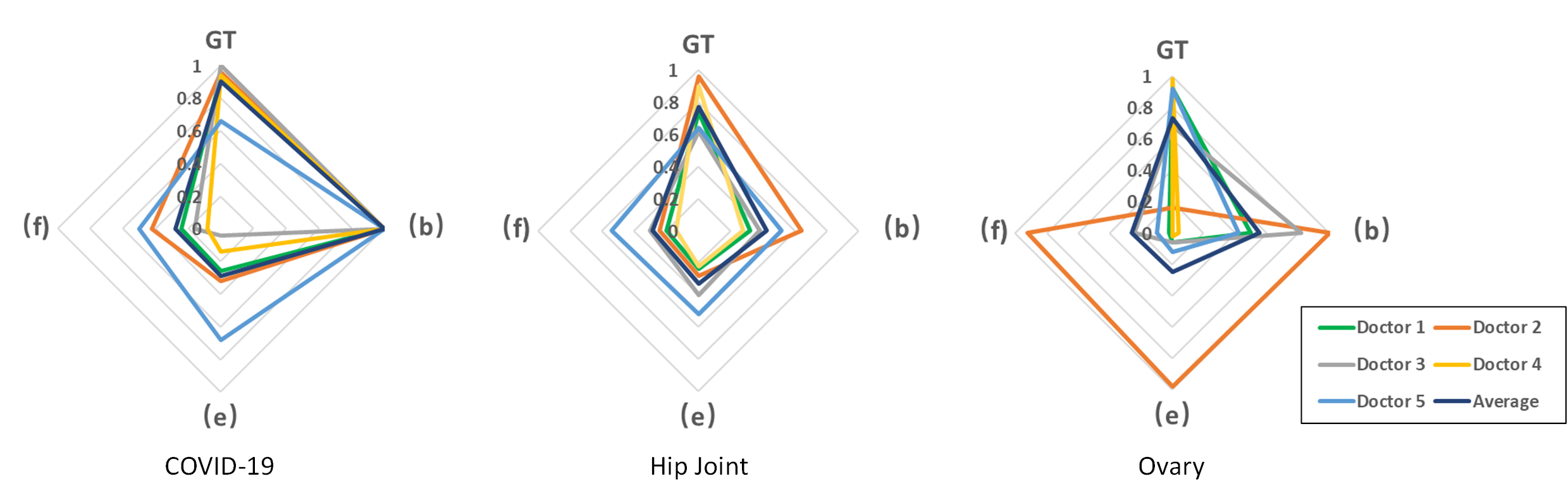}
		\caption{Radar charts showing the results of user studies. Four different types of images are used: the GT and the synthesized images by three methods (b spGAN-sketch, e spGAN, f spGAN v2). }
		\label{fig_user_study}
	\end{figure*}

	To conduct user studies \footnote{COVID-19: https://ks.wjx.top/jq/83543001.aspx} \footnote{Hip joint: https://ks.wjx.top/jq/83583052.aspx} \footnote{Ovary: https://ks.wjx.top/jq/83588557.aspx} of each dataset, we randomly chose 200 images of four different types: GT and the generated images by three methods (i.e., spGAN-sketch, spGAN, and spGAN v2 as explained in Section~\ref{Ablation}), each type with 50 images. Five doctors were asked to view these images and tell whether they were real or fake. For each of the four types, we reported the accuracy as the fraction of images being correctly classified. The results are provided in Table~\ref{table_user_study}. A lower accuracy of an image synthesis method means that the images generated by this method are more realistic, while the accuracy of GT should be close to 1. Based on the results in Table~\ref{table_user_study}, radar charts were drawn for each dataset for a more intuitive comparison, as shown in Fig.~\ref{fig_user_study}. Each vertex in the radar chart indicates the accuracy of the corresponding type of image source.
	
	\begin{table*}[!h]
	\scriptsize
	\setlength{\abovecaptionskip}{0pt}
	\setlength{\belowcaptionskip}{10pt}
	\caption{The classification accuracies of five doctors for each of the four types of images: the GT and the synthesized images by three methods (b: spGAN-sketch, e: spGAN, f: spGAN v2).}
	\label{table_user_study}
	\centering
	\begin{tabular}{c|cccc|c|cccc|c|cccc|c}
		\hline 
		\multirow{2}{*}{} & \multicolumn{5}{|c|}{COVID-19} & \multicolumn{5}{|c|}{Hip Joint} & \multicolumn{5}{|c}{Ovary} \\
		
		\cline{2-16}
		& GT & (b) & (e) & (f) & Average &
		GT & (b) & (e) & (f) & Average &
		GT & (b) & (e) & (f) & Average \\
		
		\hline
		Doctor 1 & 
		0.94 & 1.00 & 0.26 & 0.24 & 0.610 &
		0.74 & 0.32 & 0.24 & 0.20 & 0.375 & 
		0.92 & 0.50 & 0.06 & 0.02 & 0.375  \\
		
		Doctor 2 & 
		0.96 & 1.00 & 0.32 & 0.42 & 0.675 &
		0.96 & 0.64 & 0.28 & 0.24 & 0.530 &
		0.16 & 1.00 & 0.98 & 0.92 & 0.765 \\
		
		Doctor 3 & 
		1.00 & 1.00 & 0.04 & 0.16 & 0.550 & 
		0.62 & 0.38 & 0.40 & 0.30 & 0.425 &
		0.68 & 0.82 & 0.06 & 0.24 & 0.450\\
		
		Doctor 4 & 
		0.94 & 1.00 & 0.14 & 0.08 & 0.540 & 
		0.90 & 0.28 & 0.22 & 0.14 & 0.385 &
		0.98 & 0.04 & 0.02 & 0.00 & 0.260 \\
		
		Doctor 5 & 
		0.66 & 1.00 & 0.68 & 0.50 & 0.710 &
		0.64 & 0.52 & 0.52 & 0.54 & 0.555 & 
		0.92 & 0.42 & 0.12 & 0.10 & 0.390  \\
		
		\hline
		Average & 
		0.900 & 1.000 & 0.288  & \textbf{0.280} & 0.617 &
		0.772 & 0.428 & 0.332 & \textbf{0.284} & 0.454 & 
		0.732 & 0.556 & \textbf{0.248} & 0.256 & 0.448  \\	
		\hline 	
	\end{tabular}
	\end{table*}
	
	Most doctors have good ability in recognizing GT images and therefore the average accuracy (last row in Table~\ref{table_user_study}) of GT images was high.
	However, in the ovarian data set, the evaluations of Doctor 2 and other doctors are quite different. Doctor 2 tended to classify almost all images as fake whereas the other doctors tended to classify most images as real.
	
	Among different methods, the accuracy of the spGAN-sketch (i.e., the spGAN without sketch guidance) was higher especially in the COVID-19 dataset. This indicates sketch guidance plays an important role in synthesizing realistic images. Without the sketch guidance, the generated COVID-19 images present obvious checkerboard patterns (Fig.~\ref{fig_512_covid}), which makes it very easy for the doctors to distinguish them from real images.
	We also observed that the accuracy of spGAN-sketch method in the hip joint and ovary datasets was much lower than that in the COVID-19 dataset. This indicates that the sketch guidance is particularly necessary for lung US image synthesis.
	Among the three image synthesis methods, spGAN v2 achieved lower average accuracy than the other two methods in all datasets except that it achieved higher average accuracy than spGAN in the ovary dataset. This means that generally the use of the sketch guidance and feature loss can improve the quality of generated images.
	
	\begin{figure*}[!t]
		\centering
		\includegraphics[width=0.9\textwidth]{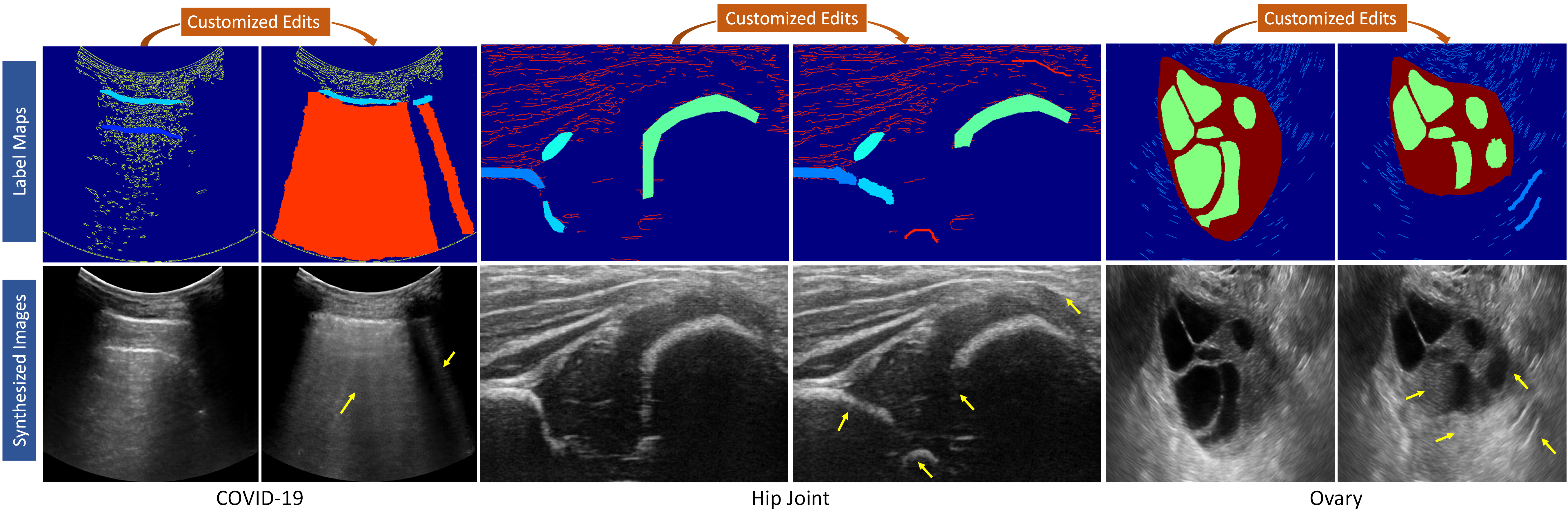}
		\caption{Examples of editable synthesis. The yellow arrows indicate the changes before and after editing.}
		\label{fig_edit}
	\end{figure*}
	
	\subsection{Editable synthesis}
	\label{Editable}
	Furthermore, we developed a platform for editable image synthesis. Via customized editing of label maps, this platform can be used as a convenient tool to simulate various, meaningful US images for training sonographers and deep neural networks.
	
	Fig.~\ref{fig_edit} shows some synthesized examples after editing the label maps, using the proposed spGAN v2 method. For the COVID-19, the artifacts, such as the pleura line, A-line, B-line, and lung consolidation, are evaluated for grading disease severity. We can simulate a more severe COVID-19 case by adding the B-line to the label map. In hip joint US images, the diagnosis of infant hip dysplasia depends on the relative position of different structures. By changing the relative position of the structures in the label map of a normal case, we can create a case of developmental dislocation of the hip with high fidelity. At last, we can easily synthesize new ovary US images by changing the size of ovary and the number of follicles in the label map.
	A video demo of the editable synthesis using our developed platform is provided in Supplementary file \emph{Editable-demo.mp4}.

	\subsection{Segmentation experiments}
	\label{Segmentation}

	To explore whether using our method for data augmentation can improve the segmentation performance, we compared the results of three settings: no augmentation (baseline), traditional augmentation (trad), and traditional plus GAN-based augmentation (trad+GAN). Furthermore, in addition to using the full training data, we also experimented with 20\% of the training data to see how these data augmentation methods performed on small datasets. We used U-Net \citep{ronneberger2015u} as the segmentation model and DICE as the performance metric.
	
	We employed online augmentation and set the possibility to 0.3 for both traditional and GAN-based augmentation. For traditional augmentation, the operations included rotation, translation, scaling, blurring, gamma transformation, and adding Gaussian noise. For GAN-based augmentation, we randomly edited the label maps by morphological operations and used spGAN v2 to synthesize images.
	
	The segmentation results are presented in Table~\ref{table_segmentation}. For the hip joint and ovary datasets, Traditional augmentation improved segmentation performance when 20\% of the training data were used but failed when the full training data were used. For the COVID-19 dataset, traditional augmentation did not improve performance no matter a portion of or the full training data were used. The reason may be that the lesion areas in COVID-19 images have rich styles and thus it is hard to gain additional information through simple operations in traditional augmentation. In contrast, our GAN-based augmentation method can provide greater variability with editable operations and therefore has great potential to improve performance. As shown in Table~\ref{table_segmentation}, our method in general not only achieved superior segmentation results when a small training set was available but also further improved performance even when the full training set was used.

	\begin{table*}[!h]
		\scriptsize
		\setlength{\abovecaptionskip}{0pt}
		\setlength{\belowcaptionskip}{10pt}
		\caption{Comparison of the segmentation performance (DICE) with and without data augmentation. 20\%: training with 20\% data.}
		\label{table_segmentation}
		\centering
		\begin{threeparttable}
			\begin{tabular}{l|llll|llll|ll}
				\hline
				
				\multirow{2}{*}{Setting} & \multicolumn{4}{c|}{COVID-19} & \multicolumn{4}{c|}{Hip Joint} &\multicolumn{2}{c}{Ovary}
				
				\\
				& Pleural Line & A-line & B-line & Consolidation & 
				Ilium & Lower Limb & Labrum & Co-junction & 
				Ovary & Follicles \\
				\hline
				Baseline (20\%) &  
				0.5068 & 0.2010 & 0.6102 & 0.0911 &
				0.8593 & 0.7953 & 0.8208 & 0.8416 &
				0.7915 & 0.7934 \\
				
				Trad (20\%) &
				\textbf{0.5076} & 0.1784 & 0.5925 & 0.1132 &
				0.8707 & 0.8020 & 0.8262 & 0.8494 &
				0.8554 & 0.8393 \\
				
				Trad+GAN (20\%) &  
				0.4974 & \textbf{0.2070} & \textbf{0.6150} & \textbf{0.1368} &
				\textbf{0.8841} & \textbf{0.8308} & \textbf{0.8323} & \textbf{0.8623} &
				\textbf{0.8740} & \textbf{0.8573} \\
				
				\hline
				Baseline &  
				0.5562 & \textbf{0.3239} & 0.7447 & 0.3657 &
				0.8866 & 0.8185 & 0.8206 & 0.8574 &
				0.9064 & 0.8863  \\
				
				Trad &
				0.5208 & 0.3121 & 0.7386 & 0.3729 &
				0.8855 & 0.8148 & 0.8013 & 0.8591 &
				0.9026 & 0.8829 \\
				
				Trad+GAN &  
				\textbf{0.5602} & 0.2933 & \textbf{0.7629} & \textbf{0.4137} &
				\textbf{0.8982} & \textbf{0.8397} & \textbf{0.8384} & \textbf{0.8657} &
				\textbf{0.9101} & \textbf{0.8876} \\			

				\hline	
			\end{tabular}	
			
		\end{threeparttable}
	\end{table*}
	
	\begin{figure*}[!t]
		\centering
		\includegraphics[width=\textwidth]{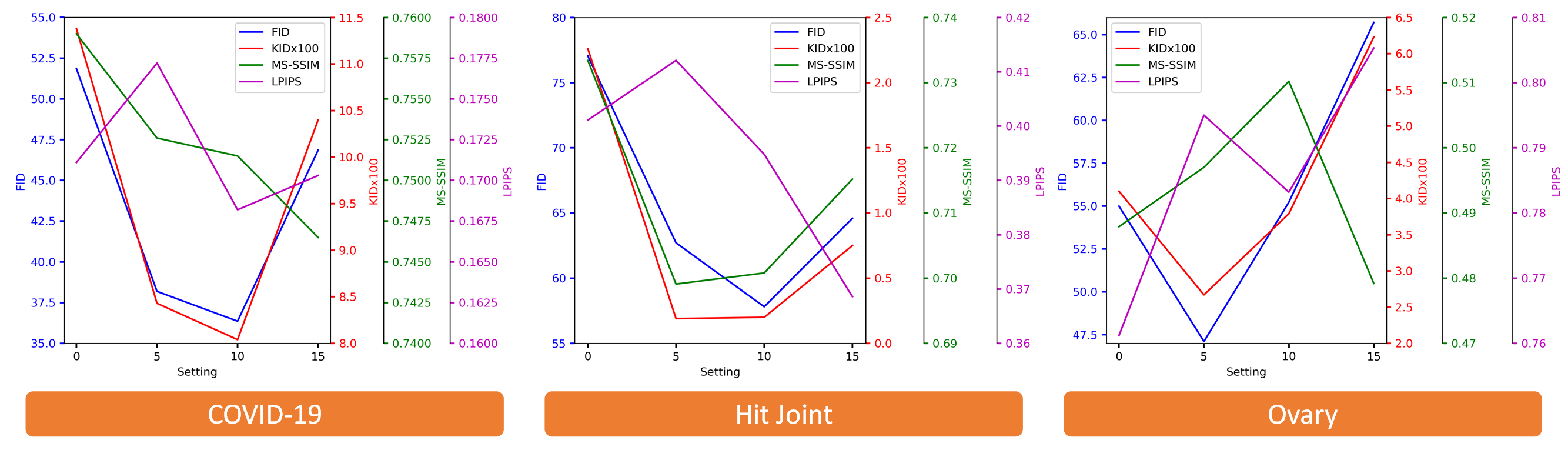}
		\caption{Effect of the weight of the feature loss in the objective function of generator. The horizontal axis denotes the weight, and the color-coded vertical axes represent four performance metrics.}
		\label{fig_weight_loss}
	\end{figure*}
	
	\begin{figure*}[!t]
		\centering
		\includegraphics[width=\textwidth]{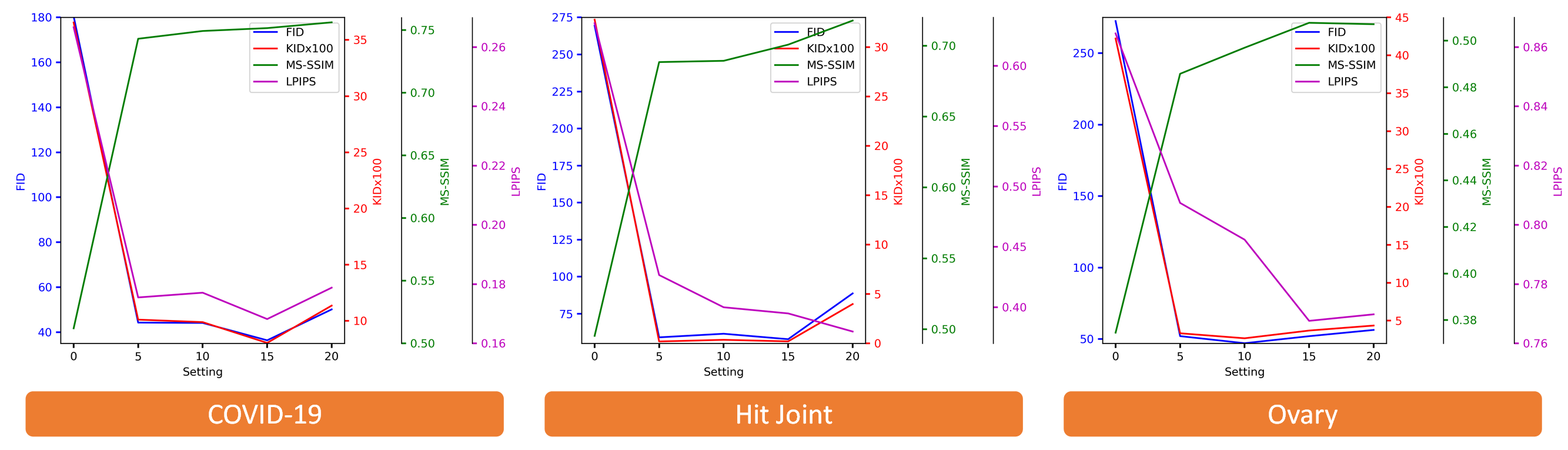}
		\caption{Effect of the number of residual blocks in generator. The horizontal axis denotes the number of residual blocks, and the color-coded vertical axes represent four performance metrics.}
		\label{fig_number_residual_block}
	\end{figure*}

	\begin{figure*}[!t]
		\centering
		\includegraphics[width=\textwidth]{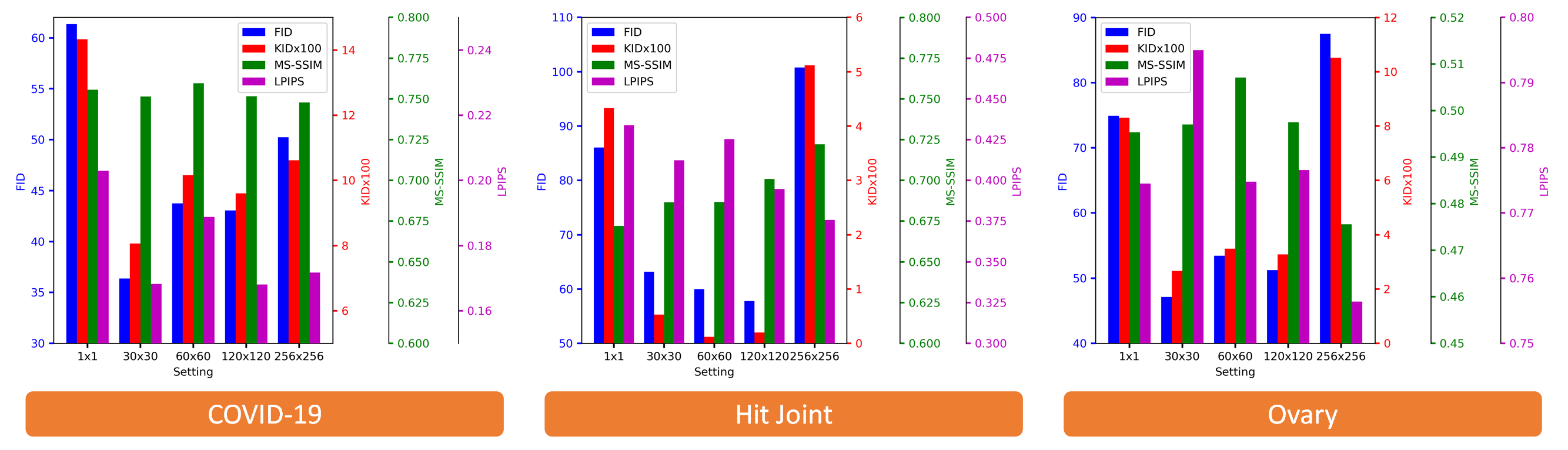}
		\caption{Effect of the output size of discriminator. The horizontal axis denotes the output size, and the color-coded vertical axes represent four performance metrics.}
		\label{fig_outputsize}
	\end{figure*}
	
	\subsection{Effect of parameters}
	\label{Parameter}

	In our proposed spGAN v2 method, there are three important parameters, i.e., the weight of feature loss term $\lambda_{2}$ in the objective function of generator, the number of residual blocks $n$ in generator, and the output size $s$ of discriminator. To investigate the effect of the three parameters on the performance of our proposed spGAN v2 method, we tested different values for parameter $\lambda_{2}$ from $\{0, 5, 10, 15\}$, parameter $n$ from $\{0, 5, 10, 15, 20\}$, and parameter $s$ from \{1$\times$1, 30$\times$30, 60$\times$60, 120$\times$120, 256$\times$256\}. Fig.~\ref{fig_weight_loss}, \ref{fig_number_residual_block}, and \ref{fig_outputsize} show the results for each of the three parameters, respectively.
	
	According to the results in Fig.~\ref{fig_weight_loss}, we chose $\lambda_{2}=10$ for the COVID-19 and hip joint datasets and $\lambda_{2}=5$ for the ovary dataset. A larger $\lambda_{2}$ means that more emphasis is put on the similarity of high-level features (feature loss) instead of the pixel-wise or structural similarity (L1-loss). Compared with the COVID-19 and the hip joint datasets, the structures (i.e., ovary and follicles) in the ovary US images have a clear boundary and regular shape. Therefore, a smaller $\lambda_{2}$ is more suitable for the ovary dataset.
	
	Based on the results in Fig.~\ref{fig_number_residual_block} and visual evaluation of the synthesized images, we set the number of residual blocks $n$ to 15, 15, and 10 for the COVID-19, hip joint, and ovary datasets, respectively. Although for the hip joint dataset, $n = 20$ achieved better performance in terms of the MS-SSIM and LPIPS metrics than $n = 15$, the visual quality of synthesized images was worse, so $n = 15$ was used for the hip joint dataset instead. Generally, more residual blocks used in the generator mean a stronger ability to learn features. However, too many residual blocks tended to degrade performance. As shown in Fig.~\ref{fig_number_residual_block}, fewer residual blocks are needed for the ovary dataset compared with the other two datasets. This is probably because the texture present in the ovary images is simpler than that in the other two types of images.
	
	The PatchGAN was used as the discriminator in our synthesis framework. Each unit of the discriminator output is like a local receptive field. A smaller output size indicates that each unit of the output represents a larger region. For instance, the output size of 1$\times$1 means that the discrimination is made from the whole image, without local information fed back to the generator. Conversely, the output size of 256$\times$256 indicates that the discrimination is made from only several pixels, neglecting the context information in surrounding region. As shown in Fig.~\ref{fig_outputsize}, the results of 1$\times$1 output size are worse in all three datasets, and too small or too large output size often degrades the performance. According to the results in Fig.~\ref{fig_outputsize} and visual quality of the generated images, we set the output size to 30$\times$30, 120$\times$120, and 30$\times$30 for the COVID-19, hip joint, and ovary datasets, respectively.

	\section{Conclusion}
	\label{Conclusion}
	In this paper, to address the challenge of lacking enough data for training sonographers and deep neural networks, we propose an image-to-image translation framework aiming at generating high-resolution and high-fidelity US images from segmentation label maps. The proposed spGAN v2 method consists of four key components: auxiliary sketch guidance, progressive growing scheme, fade-in blocks, and feature loss. Specifically, the auxiliary sketch guidance provides auxiliary information for generating realistic background texture. The progressive growing scheme containing the pre-trained model of generating low-resolution images and fade-in blocks are employed for a smooth transition from low resolution to high resolution. To further improve the quality of the generated images, the feature loss is adopted to suppress noise and deblur the images. Extensive qualitative and quantitative experiments on the COVID-19, hip joint, and ovary datasets demonstrate the effectiveness of the proposed spGAN v2 method. Another important feature of our work is that we have developed an editable image synthesis platform that can easily create various and meaningful US images by modifying segmentation label maps. Overall, our study provides a useful and convenient tool for generating high-resolution and high-fidelity US images from label maps.
	
	\section*{Acknowledgments}
	This work was supported by the grant from National Natural Science Foundation of China (No. 62171290, No. 62101343, No. 61901275), Shenzhen-Hong Kong Joint Research Program (No. SGDX20201103095613036), SZU Top Ranking Project (No. 86000000210), Shenzhen Science and Technology Innovations Committee (No. 20200812143441001), and Shenzhen University Startup Fund (No. 2019131).
	
	\bibliographystyle{model2-names}\biboptions{authoryear}
	\bibliography{refs}
	
\end{document}